\documentclass[letterpaper]{article} % DO NOT CHANGE THIS
\usepackage{aaai21}  % DO NOT CHANGE THIS
\usepackage{times}  % DO NOT CHANGE THIS
\usepackage{helvet} % DO NOT CHANGE THIS
\usepackage{courier}  % DO NOT CHANGE THIS
\usepackage[hyphens]{url}  % DO NOT CHANGE THIS
\usepackage{graphicx} % DO NOT CHANGE THIS
\usepackage{natbib}  % DO NOT CHANGE THIS AND DO NOT ADD ANY OPTIONS TO IT
\usepackage{caption} % DO NOT CHANGE THIS AND DO NOT ADD ANY OPTIONS TO IT
\usepackage{booktabs}
    
\usepackage{natbib}
\usepackage{graphicx}
\usepackage{xcolor}
\usepackage{multirow}

\usepackage[acronym]{glossaries}
\makeglossaries

\newcommand{\urlfootnote}[1]{\footnote{\url{#1}}}

\newacronym{Base}{Majority Baseline}{Majority Baseline}
\newacronym{HP}{Human Performance}{Human Performance}
\newacronym{BertBaseAURC}{ArgBERT}{ArgBERT}
\newacronym{BertBaseAURCAMSR}{PeerBERT-ArgInit}{PeerBERT-ArgInit}
\newacronym{BertBaseAMSR}{PeerBERT}{PeerBERT}
\newacronym{BertLargeAMSR}{PeerBERT-L}{PeerBERT-L}
\newacronym{ToBert}{ToBERT}{ToBERT}

\newacronym{nogi}{NO-GI}{NO-GI}
\newacronym{withgi}{ALL}{ALL}

\newenvironment{centerquote}[1][1cm]
 {\begin{center}
  \begin{tabular}{|p{\dimexpr\linewidth-#1-#1}|}
  \hline}
 {\\ \hline
  \end{tabular}
  \end{center}}

\nocopyright
\title{Argument Mining Driven Analysis of Peer-Reviews}
\author{
    Michael Fromm,\textsuperscript{\rm 1}
    Evgeniy Faerman,\textsuperscript{\rm 1}
    Max Berrendorf,\textsuperscript{\rm 1}
    Siddharth Bhargava,\textsuperscript{\rm 2}
    Ruoxia Qi,\textsuperscript{\rm 2}
    Yao Zhang,\textsuperscript{\rm 2}
    Lukas Dennert,\textsuperscript{\rm 2}
    Sophia Selle,\textsuperscript{\rm 2}
    Yang Mao,\textsuperscript{\rm 2}
    Thomas Seidl\textsuperscript{\rm 1}\\
}

\affiliations {
    \textsuperscript{\rm 1} Database Systems and Data Mining, LMU Munich, Germany \\
    \textsuperscript{\rm 2} LMU Munich, Germany \\
    \{fromm,faerman,berrendorf\}@dbs.ifi.lmu.de
}

\begin{document}

\maketitle

\begin{abstract}
% Peer reviewing is a central process for scientific discovery and it is essential for high quality in research work. 
Peer reviewing is a central process in modern research and essential for ensuring high quality and reliability of published work.
At the same time, it is a time-consuming process and increasing interest in emerging fields often results in a high review workload, especially for senior researchers in this area.
How to cope with this problem is an open question and it is vividly discussed across all major conferences.
In this work, we propose an Argument Mining based approach for the assistance of editors, meta-reviewers, and reviewers. 
We demonstrate that the decision process in the field of scientific publications is driven by arguments and automatic argument identification is helpful in various use-cases.
One of our findings is that arguments used in the peer-review process differ from arguments in other domains making the transfer of pre-trained models difficult.
Therefore, we provide the community with a new peer-review dataset from different computer science conferences with annotated arguments. 
In our extensive empirical evaluation, we show that Argument Mining can be used to efficiently extract the most relevant parts from reviews, which are paramount for the publication decision.
The process remains interpretable since the extracted arguments can be highlighted in a review without detaching them from their context.

%In our work, we develop and adapt argument mining techniques for the scientific peer review domain.
%The reviewing process in academia is the flagship domain of argumentation for decades.
%In contrast to previous work, where the focus was to analyze how reviews are structured, we set the focus on argument extraction in scientific-reviews.
%We develop and use state-of-the-art argument mining (AM) methods and test them on their generalization capacity to the scientific-review domain. 
%Our results clearly show that AM methods trained on other text types are not useable in the scientific review area.
%Therefore, we present a new dataset Argument Mining in Scientific Reviews (AMSR) which contains reviews of 3,962 recently submitted papers from major conferences in computer science. 
%A representative sample of 77 reviews got annotated on token- and sentence-level with high inter-annotator agreement. 
%Our best performing argument extraction method achieves nearly human-level-performance on the benchmark and generalizes well across different conference domains.
%We show that our AM method renders important information more precisely for different phases of the peer-reviewing process. 
%An application showcase clearly presents that the found arguments are the most important information in the reviews for the paper acceptance decision. 
%To foster reproducibility and support further research, we publicly release our code and the AMSR dataset at \todo{\url{https://github.com/TODO}}.
\end{abstract}

\section{Introduction}
Argumentation is a process of bringing together and organizing reasons to convince a reasonable critic to accept or refuse a certain standpoint \cite{van2004systematic}.
It is an essential part of each rational decision-making process and after the decision is made, argumentation is important for its explanation and justification \cite{amgoud2009using}. 
An important step in the argumentation process is the identification of arguments. 
Generally speaking, there is a difference between \emph{argumentative} and \emph{informative} content:
Argumentative content expresses evidence or reasoning used to either oppose or support a given point.
Informative parts often contain background information and describe how entities appear and act in the world.
%Recent advances in the field of \emph{Argument Mining} (AM) have demonstrated that the process of argument identification can be automated in various domains such as encyclopedic articles \citep{aharoni-etal-2014-benchmark}, student essays \citep{DBLP:conf/emnlp/StabG14}, web discourse \citep{DBLP:journals/corr/HabernalG16} or political speeches \citep{haddadan-etal-2019-yes}.

In the last years, \emph{Argument Mining} (AM) approaches have been applied in many fields and for different types of texts, such as encyclopedic articles \citep{aharoni-etal-2014-benchmark}, student essays \citep{DBLP:conf/emnlp/StabG14}, web discourse \citep{DBLP:journals/corr/HabernalG16} or political speeches \citep{haddadan-etal-2019-yes}.
%like legal reasoning \cite{Wyner2010}, \todo{decision making processes \cite{SVENSON197986}, argument search engines \citep{wachsmuth2017building, stab2018argumentext} and is even used for complex dialogical argumentation with humans \cite{slonim2018project}.}
AM techniques build the backbone of an IBM AI system \emph{Project Debater}, which has the ambitious goal to debate humans on complex topics.
This work aims to further extend the application of AM to the novel domain of scientific \emph{peer reviews}.
Peer reviewing is a cornerstone of today's academic editorial decision-making process in nearly all scientific disciplines.
The peer-reviewers, who are usually not part of the editorial team, are experts in the corresponding research field and their task is the critical evaluation of the work proposed for publication.
We argue that peer-reviewing can also be seen as an argumentation process, where the reviewers make up their minds about the examined publications and try to convince the editorial team by providing arguments in favor of or against acceptance.
%An evaluation or review is usually comprised of different parts, it often contains a summary of the work, additional background information about the topic, improvement suggestions, opinion of the reviewer whether the work should be accepted, and pro and contra arguments.
While the evaluation or review usually comprises different parts, such as a summary of the work or additional background information about the topic, the reviewers' pro and contra arguments are often the most relevant for making the final decision. 
%While most of this information is useful for the carefully considered decision
Consequently, we envision that the automatic identification of argumentative content can improve and simplify different peer-review process phases.
One possible use-case is to provide editors or meta-reviewers, co-responsible for the final decision, with an overview of arguments from all reviews and let them focus on the most relevant ones.
For instance, after reading only the highlighted arguments in Figure~\ref{fig:reviewExampleIclr20373}, it is possible to get a good idea about the paper's strong and weak points.
Another possible use-case is to support the reviewers by providing information about (missing) argumentation. 
For example, the author of the review in Figure~\ref{fig:reviewExampleIclr20373} provides a detailed description of the empirical evaluation, but it is not completely clear from the text whether the reviewer is satisfied with the proposed evaluation criteria.

\begin{figure}
    \begin{centerquote}[0cm]
    \underline{\textbf{Example Review}}\\
    %\tiny
    %\scriptsize
    \footnotesize
    \input{review_examples/iclr20_37_3}
    \end{centerquote}
    \caption{
    Example review for an ICLR'20 submission with labeling:
    Arguments in favor of acceptance are shown in \textcolor{green!50!black}{\bf bold font and green}; \textcolor{red}{\em red and italic font} denotes arguments against it.
    %Only a small fraction of the review is argumentative and the arguments are clear cut. The rest of the review mostly consists of explanations about how Mutual Information (MI) estimation works and descriptions about the experimental setup.
    }
    \label{fig:reviewExampleIclr20373}
\end{figure}

In this paper, we propose the application of AM to the domain of peer-reviewing.
To this end, we collect a new dataset containing peer-reviews from different computer science conferences. 
We define a suitable AM annotation schema and annotate the dataset.
We investigate the applicability of state-of-the-art AM techniques in an extensive empirical evaluation.
Among others, we study the transferability of models trained on data from different domains to our task and the generalization across different conferences.
Furthermore, we empirically validate our assumption about the importance of arguments for the decision-making process in academic publishing.
\section{Related Work}
\subsection{Argument Mining}
Argument Mining (AM) is the task of recognizing argument components~\cite{Palau:2009:AMD:1568234.1568246, DBLP:journals/corr/HabernalG16, stab-gurevych-2017-parsing, article2,nguyen-litman-2015-extracting} and their relations ~\cite{stab-gurevych-2017-parsing, nguyen-litman-2016-context}.
The basis of AM are argumentation schemes that define the structure of the argument components and the relations between them.
There is no universally accepted theory of argumentation~\cite{van2019handbook}, and over time, argumentation schemes of varying complexity have been suggested in the literature~\cite{Toulmin1958-TOUTUO-2,article,Freeman2011-FREASR-2,DBLP:conf/emnlp/StabG14}.
The original model by~\citet{Toulmin1958-TOUTUO-2} comprises \emph{claims} as an assertion for general acceptance, \emph{data} (also often called premises) as the source of evidence to establish the claim, a \emph{warrant} to justify the inference from a premise to a claim, \emph{backing} (facts behind the warrant), a \emph{qualifier} (degree of certainty for the inference) and \emph{rebuttals}. 
The model has often been adopted in literature and most of the time, only premises and claims are used as argument components.
However, it was observed that arguments in many text types have a more straightforward structure, e.g., models trained on a single dataset to identify \emph{claims} do not generalize well to other document types~\cite{DBLP:journals/corr/DaxenbergerEHSG17}. 
Furthermore, annotating a dataset crawled from heterogeneous text sources leads to a low agreement among annotators~\cite{DBLP:journals/corr/HabernalG16, miller-etal-2019-streamlined}. 
Also, specific argument components (backing, warrant) appearing in the Toulmin-Scheme~\cite{Toulmin1958-TOUTUO-2} are often stated implicitly~\cite{Hitchcock2003, DBLP:journals/corr/HabernalG16}. 
An argumentative scheme recently proposed by~\citet{darmstadt_1} omits these components and simply distinguishes between (\emph{supporting}/\emph{opposing}) arguments and non-argumentative text parts.
Its reasonableness is confirmed on the one hand by relatively high agreement among reviewers, and on the other hand by the model performance on texts from heterogeneous sources, see e.g.~\cite{DBLP:conf/webi/FrommF019}.
Furthermore, it was observed that the distinction between \emph{supporting} and \emph{opposing} arguments is more challenging than the distinction between argumentative and non-argumentative parts~\citep{trautmann2020relational, trautmann2020fine, DBLP:conf/webi/FrommF019}.
%More recently an argument scheme was introduced that is more simple and captures the argumentativeness (supporting, opposing or non-related regarding a topic) in sentences \citep{darmstadt_1}. They proved that arguments can be annotated with sufficient agreement in heterogeneous sources. 

The development of models for the identification of argument components according to an argumentative scheme is similar to other NLP disciplines.
Previous approaches rely on feature engineering~\citep{DBLP:journals/corr/HabernalG16, lawrence-reed-2015-combining, stab-gurevych-2014-annotating}, more recent methods apply neural networks models. 
\citet{guggilla-etal-2016-cnn} were the first to apply recurrent neural networks for AM.
The state-of-the-art performance in AM is achieved with pre-trained transformer-based architectures~\cite{DBLP:conf/webi/FrommF019, trautmann2020fine,reimers-etal-2019-classification}.

A popular real-life application of AM techniques are argument search engines such as \emph{argumenText}\urlfootnote{www.argumentsearch.com}~\citep{stab2018argumentext} and \emph{args}\urlfootnote{www.args.me}~\citep{wachsmuth-etal-2017-building} which allow argument retrieval according to a user-defined topic.
AM is applied in the preprocessing step, where arguments are extracted from documents before they are indexed by a search engine. 

\subsection{Application of NLP for Peer-Reviewing Process}
So far, AM for scientific peer-reviews has received little attention. 
\citet{hua-etal-2019-argument-mining} introduce a dataset with propositions in scientific reviews. 
The annotation schema is comprised of components that often appear in reviews such as \emph{requests, facts, evaluations} or \emph{quotes}.
The dataset is annotated on a sentence level and the main focus is to study the usage of different propositions across venues. 
In our application, we are interested in arguments directly affecting the decision process, and therefore, the stance of the argument bears essential information. 
Since this information is missing in~\citet{hua-etal-2019-argument-mining}, this annotation schema is not suitable for our application.
%the interplay between reviewers and meta-reviewers and therefore this annotation schema is not suitable for us. The argumentation schema does not contain information regarding the stance of the propositions, which are needed in our application.
Closely related is \citet{xiao2020detecting} work, where the goal is to automatically detect the problem description in peer-reviews.
However, although the problems can also be considered opposing arguments, it is crucial to consider both positive and negative arguments for our application.

Other related works deal with different aspects of the peer-reviewing process. 
In~\citet{plank2019citetracked}, the authors introduce a dataset with scientific reviews and analyze it based on the title, abstract, and review text on how well the citation impact of a paper can be predicted.
\citet{DBLP:journals/corr/abs-1903-11367} study the effect of author replies in the rebuttal phase. 
\emph{Argumentative zoning}~\cite{teufel-etal-2009-towards} analyzes the rhetorical and argumentative structure of scientific papers with intending to convince reviewers that the knowledge claim of the paper is valid.

\section{Dataset}
We use the OpenReview\urlfootnote{https://openreview.net/} platform and the OpenReview-Crawler\urlfootnote{https://openreview-py.readthedocs.io/en/latest/getting_data.html} to retrieve peer-reviews. We collect all reviews from six computer-science conferences listed in Table~\ref{tab:conferences}.
The annotated dataset \urlfootnote{https://zenodo.org/record/4314390} and the code \urlfootnote{https://github.com/fromm-m/aaai2021-am-peer-reviews} is available.

There, we additionally provide basic statistics about conferences and collected reviews.

\begin{table*}
    \centering
    \begin{tabular}{lrrcc}
    Conference & Number of Papers & Number of Reviews & Acceptance rate & avg words\\
    \toprule
    ICLR'19 & 1,419 & 4,332 & 35 \% & 403	 \\
    ICLR'20 & 2,213 & 6,722 & 27 \% & 409 \\
    MIDL'19 & 59 & 178 & 80 \% & 362\\
    MIDL'20 & 144 & 544 & 55 \% & 255\\
    NeuroAI'19 & 62 & 174 & 68 \% & 305\\
    GI'20 & 65 & 174 & 82 \% & 507\\ 
    \bottomrule
    Total & 3,962 & 12,135 & - & 368\\
    \end{tabular}
    \caption{
    Dataset statistics
    }
    \label{tab:conferences}
\end{table*}

\subsection{Preprocessing}
%Each paper contains information of the reviews, possible comments of the authors, decisions and the meta-review. 
In a first preprocessing step, we replace URLs, escape sequences, encapsulated mathematical formulas, Unicode symbols and markdown with a corresponding type placeholder token respectively, e.g. \texttt{<URL>} for URLs.
Furthermore, we remove multiple consecutive whitespaces and split review texts into sentences using the \texttt{PunktSentenceTokenizer} from NLTK.\urlfootnote{https://www.nltk.org/} 
To further improve the sentence splitting results, we provide the tokenizer with a set of idioms and abbreviations commonly used in scientific texts to avoid sentence splitting in the middle or after them.\footnote{The manually defined set contains e.g. "e.g", "i.e.", "et al.", "Fig.", etc.} 
Finally, we remove all sentences with less than three tokens and go through the dataset manually and remove non-interpretable sentences.

From 12,135 collected reviews, we sample 77 for the annotation.
To this end, we first sample a conference uniformly at random and then a review from the conference.\footnote{We end up with 15 reviews for iclr20, 14 reviews for iclr19 and 12 per each other conference} 
We use stratified sampling to ensure that sampled reviews reflect the following three characteristics of original review distribution for each conference:
%We samples 77 reviews uniformly across all conferences and mirror the original distribution of the whole dataset () regarding the following properties.
Review-Rating (1-4), Paper-Decision (acceptance / rejection), and Review-Length.
% \begin{itemize}
%     \item Review-Rating (1-4)
%     \item Paper-Decision (acceptance / rejection)
%     \item Review-Length 
% \end{itemize}
%The statistics of the annotated dataset can be seen in the Table \ref{tab:annotations}.

%additionally provide it with a set of common abbreviations in scientific texts.
%Moreover, the reviews often contain text that does not have the form of complete, grammatically correct sentences, which is especially the case for references.
%Therefore, we remove sentences with less than three tokens. Additionally, we manually removed some of these incomplete sentences.

\subsection{Annotation}
% rhetoric questions, reviewer waren sich uneinig was es für komponenten sind
%Van Eemeren et al. \citep{book} admit in their recent survey of the field that there is no unitary theory of argumentation that is universally accepted. The annotation scheme is therefore use-case specific. A annotation study on the \citep{Toulmin1958-TOUTUO-2}
%schema showed that argument components such as qualifier, which states the degree of cogency, warrants as logical explanations, rebuttals as statements that attack the claim, refutation which are used to attack the rebuttal, are all mostly absent in the data. Furthermore we could not achieve a reliable inter-annotator-agreement on these components. 
%The presence of these components are crucial for argumentation, however most of the times the parts are implicit which is rather usual in ordinary argumentative discourse and is known as enthymematic argumentation \citep{argumentation}. Similar results were found for heterogeneous text domains across the web \citep{DBLP:journals/corr/HabernalG16}.
%In a small study with 30 reviews we've tried to do annotations with different argumentative components such as \emph{claim, premise, warrant, backing, qualifier} and \emph{rebuttal}. However, while we obtain relatively high agreement regarding  
%In a small study  30 reviews were analyzed by a graduate level computer science researcher for the occurrence of different argumentative components, such as \emph{claim, premise, warrant, backing} or \emph{qualifier}.
\paragraph{Scheme}
We use a simple argumentation scheme proposed in
\citet{darmstadt_1}, which distinguishes between non-arguments, supporting arguments and attacking arguments, which we denote as \texttt{NON/PRO/CON} accordingly. 
While this simple scheme grasps argumentative context, the annotation is easier since annotators are not required to consider complex relationships between argumentative components.
Furthermore, it is also flexible enough to capture argumentative parts that are not attributable to the single argument type.
For instance, in our dataset, we often observe rhetorical questions that criticize the paper's vagueness under review.
The annotation scheme can also be interpreted as a flat version of the \emph{claim-premise} model:
There is a single \emph{claim}, \emph{"The paper should be accepted"}, and arguments are premises that either attack or support the claim. 
%In our dataset an argument always contains a span of text expressing evidence or reasoning either for (supporting) or against (opposing) the acceptance of a paper. 
%and to assess the reasonableness of the selected approach we rely on the agreement among annotators. 
%The annotation guidelines will be released. For the annotation process we used the web-based annotation tool WebAnno.\urlfootnote{https://webanno.github.io/webanno/}

\paragraph{Annotation Process}
\begin{table*}
    \centering
    \begin{tabular}{lrrrr}
        & PRO & CON & NON & Total\\
        \toprule
        number of tokens & 3,259 (12\%) & 10,559 (34\%) & 14,684 (54\%) & 28,502 \\
        number of sentences & 203 (14\%) & 640 (46\%) & 558 (40\%) & 1,401 \\
    \end{tabular}
    \caption{
    The table shows the distribution of the classes in the datasets. 
    The distribution of the labels in the token-level dataset is skewed towards \texttt{NON}, and in the sentence-level dataset towards \texttt{CON}.
    }
    \label{tab:annotations}
\end{table*}
In total, we have seven annotators, all of whom are graduate-level computer science students. 
The annotation is made token-wise and when presented a review, an annotator chooses argumentative text spans and assigns labels with the argument type to it.
The document parts which are not explicitly annotated are considered to be non-argumentative. 
We refer to this annotation as \emph{token-level} annotation.

Each review is randomly assigned to three different annotators.
We resolve situations when a token is assigned with different labels by different annotators with a majority vote.
In case a token is assigned with three different labels, we ask a independent fourth annotator who did not previously annotate the review to make the final annotation decision.
%breaks the tie
%The individual annotations by the 3 different annotators per review are combined into a unified one by majority vote.
%In case all three annotators labeled differently, i.e. \texttt{PRO}/\texttt{CON}/\texttt{NON}, there is no majority. 
%In that case, an independent fourth annotator who did not previously annotate the review breaks the tie and decides alone on the label of the segment.
%We create two versions of the dataset for the experiments on sentence level and on segments respectively. 
%\todo{Both contain a column with the sentences and a column with the sentence ID which is composed of the conference name, the paper number, the review number for this paper and the sentence number in this review. The dataset for the experiments on sentence level includes a column with the stance of the complete sentence. The dataset for the segment-based experiments contains a column which lists all the segments that are longer than one character with the beginning, length and stance of the segment.} 
%\todo{Both contain a sentence ID and the sentences. One dataset includes the stance of the complete sentence. The other one lists the segments with the beginning, length and stance of the segments for each sentence.}

%sentence position
%\paragraph{Sentence-Level Annotations}
%The sentences are annotated on token level.
%To obtain sentence-level annotations we apply an approach similar to \cite{trautmann2020fine}.
To obtain \emph{sentence-level} annotations from annotated tokens, we mainly follow the procedure described in \citet{trautmann2020fine}.
Sentences without argumentative tokens are annotated with the label \texttt{NON}.
For sentences containing argumentative tokens, we count the number of argumentative segments, which overlap with it.
An argumentative segment is comprised of a sequence of tokens with the same argumentative label without interruption.
The sentence is assigned with the label of the majority of segments.
If the number of segments with both labels is the same, we count the number of tokens with argumentative labels and assign the most frequent token label.
%If the sentence does not contain positive or negative arguments it is labeled \texttt{NON}, meaning non-argumentative. 
%If it only contains positive segments and non-argumentative parts it is labeled as \texttt{PRO}, with just negative segments and non-argumentative parts it is labeled as \texttt{CON}.
%Sentences that contain both positive and negative segments are labeled with the more frequent label.
%In cases where both occur equally often, we differ from \cite{trautmann2020fine} and calculate the sum of the lengths of all positive and negative segments.
%The sentence gets labeled with the label of the longer one. As a result we get a sentence-level-dataset with 1,401 sentences. (see Figure \ref{tab:annotations})
As a result, we get 28,502 annotated tokens and 1,401 sentences.
Table~\ref{tab:annotations} presents the resulting class distribution.

\subsection{Agreement}
The agreement among annotators is an important criterion for the reliability of the annotation.
Since our annotations are done on a token level and we have more than two annotators per review, we use the Krippendorff's alpha~\cite{krippendorff2016reliability} family of measures to assess the annotation quality. 
Each annotation can be seen as a set of annotated segments $(start, stop, label)$, where $start$ and $stop$ denote the segment's bounds and $label$ its class.
We include all three classes for the computation of agreement.\footnote{The score also accounts for imbalanced classes, see e.g. ~\cite{artstein2008inter}.} 
Krippendorff's alpha now considers all pairs of overlapping segments and compares the expected and the observed disagreements in the annotations. 
%It supports multiple annotators and fixed boundaries for annotated units aren't required, therefore Krippendorff's alpha meets both our requirements. 
For better comparability we follow recent related work~\cite{trautmann2020fine} and compute the following two variants:
%To assess the quality of the annotations we use Krippendorff's Alpha~\cite{krippendorff2016reliability}, since it measures the reliability of unitizing textual continua, i.e. annotated units without fixed boundaries.
%It is also suitable for multiple annotators. 
%We compute different coefficients of Krippendorff's family of alpha.
$_{cu}\alpha$ only considers the agreement in the label, while $_{u}\alpha$ additionally takes the length of the overlap into account.
For both variants, the perfect agreement corresponds to the value of 1, the score for a random agreement is zero and negative values are possible if the agreement is worse than random.
% \begin{itemize}
%     \item $_{cu}\alpha$ % ohne länge
%     takes only overlapping into account and measures the agreement in overlapping segment parts
%     \item $_{u}\alpha$ % mit länge
%     additionally takes the level of overlapping into account.
%   %\item $_{u}\alpha$ applies to all segments, including the annotated units and the gaps between them.
%   %Since we regard non-annotated parts as non-argumentative (sub)sentences, which will also be used as input of our model, it is necessary to take them into consideration. 
%   %\item $_{cu}\alpha$ focuses only on annotated units and ignores the gaps.
%   %It indicates the level of confidence in annotating argumentative segments.
% \end{itemize}
%\todo{Krippendorff's alpha penalizes variables with strongly skewed distributions.
%It compares observed disagreement with the expected disagreement, where the expected one is strongly influenced by the ratio of values.
%If the number of annotations with one label is much higher than other ones, the expected disagreement could be very low, resulting in a low alpha score, potentially even negative.}
For our annotation, we obtain $_{u}\alpha = 0.568$ and $_{cu}\alpha = 0.861$, which is comparable to related work \citep{trautmann2020fine}.

Another possibility to assess the agreement is to compute the Macro $F_1$ metric for individual annotators. In terms of the Macro $F_1$ score, the quality of our annotations is better than of comparable  datasets \citep{trautmann2020fine, reimers-etal-2019-classification}, see Human Performance in  Table~\ref{tab:results}.
Thus, we conclude that our annotation is reliable for further experiments.
%The value of $_{cu}\alpha$, which only uses the annotated parts is significantly larger than that of $_{u}\alpha$, which is also applied on non-annotated units.
%There is far less disagreement on positive vs. negative arguments.
%\begin{table*}
%    \centering
%    \caption{Agreement comparison}
%        \label{tab:agreement}
%    \begin{tabular}{llll}
%     work & text gerne & labels & score \\
%    \toprule
%    AURC \cite{trautmann2020fine} & web texts & PRO/CON/NON & %$_{u}\alpha(merged) = 0.61 $ \\
%    NEWS \cite{eckle-kohler-etal-2015-role} & educational system news & Claim/Premise/NON & $_{u}\alpha(Prem/Cl) = 0.417 $\\
%    Reviews \cite{inproceedings} & hotel reviews & Major Claim / Claim / Attack / Support & $_{cu}\alpha = 0.5 - 0.9$ \\
%    Essays \cite{stab-gurevych-2014-annotating} & student essays & %support/attack & $_{cu}\alpha(sup.) = 0.8120, _{cu}\alpha(at.) = 0.8066$ \\
%    \end{tabular}
%\end{table*}
%Our $_{u}\alpha$ of 0.568 is comparable to \cite{trautmann2020fine}, but higher than \cite{eckle-kohler-etal-2015-role}, which also contains two tags and gaps. \cite{inproceedings} and \cite{stab-gurevych-2014-annotating} use another score, which is similar to our $_{cu}\alpha$ but uses a different distance function. Our score is higher than both of them, even when they calculate $_{cu}\alpha$ for each label separately. Since our agreement scores are in line with related studies \cite{trautmann2020fine, eckle-kohler-etal-2015-role} our annotation is reliable for further experiments.

\section{Experimental Setup}
In the following, we discuss our experimental setup.
The description applies for both token-level and sentence-level evaluation unless noted otherwise. 

\subsubsection{Problem Setting}
%Our approach targets the problem of argument classification in scientific reviews.
Our goal is to identify \emph{supporting} and \emph{opposing} arguments in scientific peer-reviews and separate them from non-argumentative text.
To get a detailed analysis of the models' performance and possible bottlenecks, we first decouple the problem of \emph{argument identification} from \emph{stance detection} and solve them separately. 
Afterward, we jointly solve both problems by a single model and obtain a model performance for our desired application.
Therefore, we define the following tasks:
\begin{enumerate}
    \item Argumentation Detection: 
    A binary classification of whether a text span is an argument.
    The classes are denoted by \texttt{ARG} and \texttt{NON}, where \texttt{ARG} is the union of \texttt{PRO} and \texttt{CON} classes.
    \item Stance Detection: 
    A binary classification whether an argumentative text span is supporting or opposing the paper acceptance.
    The model is trained and evaluated only on argumentative \texttt{PRO} and \texttt{CON} text spans.
    \item Joint Detection: 
    A multi-class classification between the classes \texttt{PRO}, \texttt{CON} and \texttt{NON}, i.e. the combination of argumentation and stance detection.
\end{enumerate}
%This evaluation scheme is also frequently used in related work~\citep{DBLP:conf/webi/FrommF019, darmstadt_1, trautmann2020fine}.
%We evaluate all three tasks for both sentence-level and a token-leven annotations.

%Furthermore, we differentiate between sentence-wise and a token-wise classification for each of the tasks.

\subsection{Evaluation}
We split our dataset sentence-wise 7:1:2 into training, validation and test sets stratified by class, i.e. keeping the same ratio among classes in all three subsets.
The validation set is used for hyperparameter optimization and early stopping, whereas the test set is only used to evaluate the final model performance reported in the result section. 
We report the macro $F_{1}$ score. 
The $F_{1}$ is defined as the harmonic mean of precision and recall and Macro $F_{1}$ is the mean over the class-individual scores. 
Since Macro $F_{1}$ weights classes equally independently of class' size, it is insensitive to the class imbalance problem.
%We decided to use the macro F$_{1}$ score because it is insensitive to the imbalance of the classes (see Table \ref{tab:annotations}) and better reflects the performance of our methods by giving the \texttt{PRO} and \texttt{CON} classes a higher importance.
We train each model ten times with different random seeds and report the mean performance.\footnote{To avoid the clutter, we provide the variance across the different runs in the appendix}
%To avoid bad initialization and local minimas we train each model ten times and report the mean Macro-F1 score based on the test-set
To check the significance of our results, we use a two-sided t-test with a significance level 1\%.

\subsection{Methods}
%In our work, we evaluate and extend the AM models that perform well on sentence-level\cite{DBLP:conf/webi/FrommF019, reimers-etal-2019-classification, trautmann2020fine} and token-level \cite{trautmann2020fine}.
Since transfer learning achieves state-of-the-art results for AM on different datasets~\citep{reimers-etal-2019-classification, DBLP:conf/webi/FrommF019, trautmann2020fine} we also apply it for our task.
We employ a transformer~\cite{vaswani2017attention} based BERT model \cite{devlin2018bert} with fine-tuning on different datasets.
We include the following model variants in our evaluation:
\begin{description}
    \item[\acrshort{Base}]
    The majority baseline labels the instances with the most frequent class.
    %In the stance-detection task, the majority class is \texttt{CON} in both annotation settings, in the argument-detection task the majority class is \texttt{NON} for the token-level and \texttt{ARG} for the sentence-level. In the combined setting the majority class is \texttt{NON} for the token-level and \texttt{CON} for the sentence-level.
    %\item[BERT]
    %We include pretrained models of BERT \cite{devlin2018bert} as a recent model that achieved state-of-the-art results on many tasks including sequence classification and sequence labeling.
    %We further distinguish our %self-attention
    %models between 
    \item [\acrshort{BertBaseAURC}] 
    To assess the new dataset necessity, we evaluate the zero-shot learning performance of a BERT model fine-tuned on another AM dataset annotated on token and sentence level with the same scheme~\citep{trautmann2020fine}.
    The other dataset comprises heterogeneous data found on the internet, and therefore, the resulting model is supposed to be universally applicable.
    %a finetuned model on the  "Fine-Grained Argument Unit Recognition and Classification" (AURC)  \urlfootnote{https://github.com/trtm/AURC} to test if such a model is applicable on our dataset.
    \item [\acrshort{BertBaseAURCAMSR}] 
    We initialize the model with the weights of \acrshort{BertBaseAURC} and additionally fine-tune it on our new dataset.
    We hypothesize that the model can take advantage of the argumentative structure learned on another dataset.
    %, a model first trained on AURC and afterwards on our dataset (AMSR) for the evaluation if pretraining on a similar argument-mining dataset is beneficial
    \item[\acrshort{BertBaseAMSR}] Smaller BERT model with 110M parameters fine-tuned on our dataset (based on \texttt{bert-base-cased}).
    \item[\acrshort{BertLargeAMSR}] Larger BERT model with 340M parameters fine-tuned on our dataset (based on \texttt{bert-large-cased}).
    %and \emph{\BertLargeAMSR} are directly fine-tuned on our dataset (AMSR) and only differ in the parameter size of the model.
    %\item[\ToBert]
    %ToBERT \cite{pappagari2019hierarchical} extends the fine-tuning procedure for longer inputs by chunking the reviews into multiple segments. 
    %The individual segments are used for the finetuning of a pre-trained BERT to predict paper acceptance/rejection. 
    %A second transformer model on-top is used to combine the individual review representations into a collective representation and obtain the final classification decision. 
    \item[\acrshort{HP}] 
    An interesting experiment for assessing the applicability of the proposed solution is the comparison with the human performance on the task.
    To compute the human performance, we treat each annotator analogously to the model.
    Therefore, we compare labels produced by each annotator to the final annotations and compute the Macro F$_{1}$ score. 
    The reported score is the mean among scores of all annotators.\footnote{The resulting score should be seen as the upper bound for human performance since we use the same annotations for ground-truth.}
    %We consider the human performance as the mean across all seven annotators in comparison to the majority vote (our ground-truth).
\end{description}

\subsection{Training}
We use a weighted cross-entropy loss to tackle the class imbalance problem, where the weight is given as the reciprocal of the number of samples of this class.
The class weights are defined individually for each task and dataset.
The models are trained using either \texttt{bert-base-cased} or \texttt{bert-large-cased}, with training batch size 100 for bert-base and 32 for bert-large. 
We use the AdamW optimizer with a learning rate of $10^{-5}$ for all models and  early stopping with a patience of 3. 
%, and maximum sequence length of 64 for the AURC data and 85 for the AMSR data. Each experiment was repeated for ten different seeds.

\section{Results}
In this section, we present the results of our experiments, which we have designed to answer the following research questions:
\begin{enumerate}
    \item How well does the automatic mining of arguments work for peer-reviews?
    \item Can we transfer knowledge from pre-existing annotated argumentation datasets? 
    \item How well does the approach generalize across different conferences?
    \item How relevant are arguments in the decision making process for scientific publications?
    %Is our method able to extract arguments which are able to support the paper-acceptance-decision?
\end{enumerate}

\subsection{Automatic Mining of Arguments}
\begin{table*}
    \centering
    \begin{tabular}{lrrrrrr}
    \toprule
    Detection & \multicolumn{2}{c}{Argument} & \multicolumn{2}{c}{Stance} & \multicolumn{2}{c}{Joint} \\
    Level & Sentence & Token & Sentence & Token & Sentence & Token \\
    \midrule
    % {\scriptsize$\pm .000$}
    \acrshort{Base} & 0.351 & 0.350 & 0.423 & 0.434 & 0.234 & 0.233\\
    \midrule
    \acrshort{BertBaseAURC} & 0.316 & 0.353 & 0.719 & 0.644 & 0.203 & 0.241\\
    \acrshort{BertBaseAURCAMSR} & 0.718 & 0.877 & 0.852 & 0.862 & 0.734 & 0.796 \\
    \acrshort{BertBaseAMSR} & \textbf{0.789} & 0.896  & 0.893  & 0.849  & 0.728  & 0.808 \\
    \acrshort{BertLargeAMSR} & 0.763  & \textbf{0.900} & \textbf{0.936}  & \textbf{0.930}  & \textbf{0.757} & \textbf{0.839} \\
    \midrule
    \acrshort{HP} & 0.885 & 0.873 & 0.978 & 0.980 & 0.881 & 0.860\\
    \bottomrule
    \end{tabular}
    \caption{%
        Overview of the results for different Argument Mining tasks on token and sentence level.
        We show results in terms of Macro $F_1$ for different BERT model variants, as well as the majority baseline and human performance estimate.
        In bold font, we highlight the best performance of our models per task and level.
    }
    \label{tab:results}
\end{table*}

\begin{table}
    \centering
    \begin{tabular}{lrrrrrr}
    \toprule
    Detection & \multicolumn{2}{c}{Argument} & \multicolumn{2}{c}{Joint} \\
    Level & Sentence & Token & Sentence & Token \\
    \midrule
    % {\scriptsize$\pm .000$}
    UKP  & 0.810 & - & 0.690 & - \\
    AURC & - & 0.782 & 0.725 & 0.743 \\
    Ours & 0.789 & 0.900 & 0.757 & 0.839 \\
    \bottomrule
    \end{tabular}
    \caption{%
    Comparison of maximum Macro $F_{1}$ values obtained for different datasets from literature, UKP~\citep{darmstadt_1, DBLP:conf/webi/FrommF019} and AURC~\citep{trautmann2020fine}.
    }
    \label{tab:other_datasets}
\end{table}
The results for the three AM tasks and all methods are summarized in Table~\ref{tab:results}. 
Our most important observation is that automatic argument extraction performs close to human performance and can be relied upon in the peer-review domain.
%The most important observation from these results is that our best \acrshort{BertLargeAMSR} model performs close to human performance and, therefore, can be used to extract arguments.}
% for argument detection we are better than human-performance now
Surprisingly, the detection of the stance in the peer-review domain appears to be considerably easier than identifying arguments. 
For other datasets annotated with the same scheme, we observe an inverse effect, see Table~\ref{tab:other_datasets}.
Although there is no explicit stance detection experiment in the other works, we can infer it from the inferior results of joint detection compared against the argument detection results.

When comparing our results to other datasets on the token level, we observe that our results are substantially better, with a difference of about 10 \% points.
A reason might be that we operate on a single domain while other datasets contain heterogeneous documents covering multiple domains. 
However, we observe a significant performance difference when comparing our results on sentence and token level.
To identify the reasons, we analyze the label ambiguity within sentences in our dataset. 
We found out that 22\% of sentences for the argumentation detection task and 23\% of those for the stance detection task contain tokens annotated with both classes. 
Therefore, we conclude that while it is still possible to achieve acceptable performance on the sentence level, the difference to the token level is more evident in our dataset.

% sentence-setup
% ARG_NONARG pvalue=0.00015811468190264427 -> ohne  ArgInit signifikant besser 
% POS_NEG pvalue=1.3474964207533462e-06 -> ohne ArgInit signifkant besser 
% POS_NEG_NON pvalue=1 -> gleiche distribution
%
% token-setup
% ARG_NONARG pvalue=0.16425971300412304 -> gleiche distribution
% POS_NEG pvalue=0.04287892481736107 -> gleiche distribution
% POS_NEG_NON pvalue=0.003907587547456759 -> ohne ArgInit signifikant besser

Finally, the experiment regarding knowledge transfer from another AM dataset reveals transfer difficulties.
The zero-shot performance is better than the majority vote only on the simpler stance detection task, but it is clearly outperformed by the models directly trained on our dataset. 
The additional intermediate fine-tuning step on the other AM dataset does not bring significant improvement either compared to directly fine-tuning on our dataset, cf. \acrshort{BertBaseAMSR}. %  (p value $<$\todo{x}) 

\paragraph{Training set size}
\begin{figure}
\includegraphics[width=\linewidth]{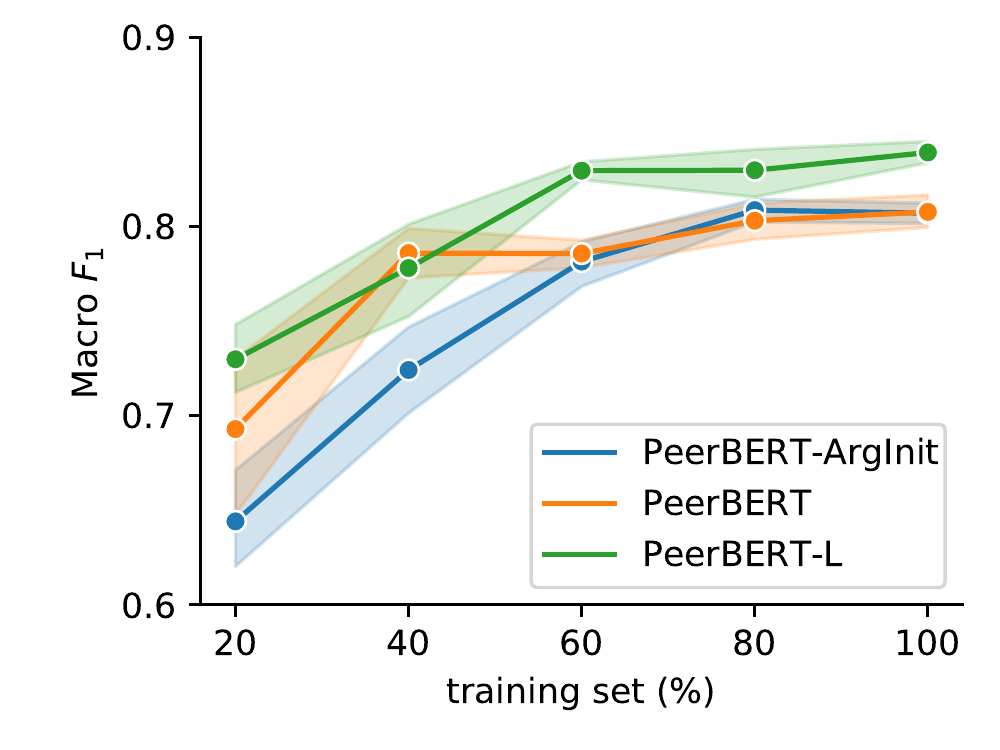}
    \caption{
    The Macro-$F_{1}$ evaluated on the task of joint prediction on the token level.
    The shaded areas indicate confidence intervals across ten runs with different random seeds.
    %The areas indicate 95\% confidence intervals obtained from bootstrapping the measurements.
    }
    \label{fig:training_sizes}
\end{figure}
Figure~\ref{fig:training_sizes} presents the model performance for different training set sizes.
We can observe that pretraining on the other AM dataset does not help, even if the training set is small.
%also if the training set is small, but it is advantageous to train a smaller model.
The performance saturates when about 60\% of the training set is used.
Therefore, we conclude that we have collected enough annotations.
%We also experiment with training data sizes. This experiment was conducted to see if enough data has been used by us to perform the requested task. From the Figure~\ref{fig:training_sizes}, showing the F$_{1}$-values at different training set sizes for Both task at token level setup, we see that for our three training methods (lines 2 - 4, Table~\ref{tab:results}), the performance plateaus when using 60\% of our data set, indicating that we have enough data for performing the desired tasks. 
Similar behavior has been observed for the other tasks at both sentence and token level.
% This paragraph speaks about topic information and can be removed if not required as we anyway are not showing any results for it. 
%\textbf{Topic.} As the AURC data set had 8 topics distributed across it, we decided to conduct experiments as well incorporating topic information into our AMSR data set. We chose to keep the topic field as "paper quality" and ran tests on our models to see if there is any improvement to the performance of the model. However, we observed that no significant improvement was achieved when the topic information was incorporated and we decided to continue our experiments without any topic information incorporated into our data and model.

%\subsection{Generalization across conferences}
%\begin{table*}
%    \centering
%    \begin{tabular}{lrrrrrr}
%    \toprule
%    Detection & \multicolumn{2}{c}{Argument} & \multicolumn{2}{c}{Stance} & \multicolumn{2}{c}{Joint} \\
%    Level & Sentence & Token & Sentence & Token & Sentence & Token \\
%    \midrule
%    \acrshort{withgi} & 0.750 & 0.890 & 0.902 & 0.942 & 0.738 & 0.831\\
%    \acrshort{nogi} & 0.718 & 0.879 & 0.901 & 0.911 & 0.718 & 0.823 \\
%    \midrule
%    \end{tabular}
%    \caption{%
%    Comparison of Macro $F_{1}$ values for sentences from GI-20 reviews, when training with/without sentences from reviews from GI-20.
%    }
%    \label{tab:gen_results}
%\end{table*}

\begin{table}
    \centering
    \begin{tabular}{lrrr}
    \toprule
    Detection & Argument & Joint \\
    \midrule
    \acrshort{withgi} & 0.891 & 0.823 \\
    \acrshort{nogi} & 0.873 & 0.791  \\
    \midrule
    \end{tabular}
    \caption{%
    Comparison of Macro $F_{1}$ values for sentences from GI-20 reviews, when training with/without sentences from reviews from GI-20. All tasks are done on token-level.}
    \label{tab:gen_results}
\end{table}

\subsection{Generalization across conferences}
In this section, we study the model's generalization to peer-reviews for papers from other (sub)domains.
To this end, we reduce the test set to only contain reviews from the GI'20 conference.
The focus of the GI'20 conference is Computer Graphics and Human-Computer Interaction, while the other conferences are focused on Representation Learning, AI and Medical Imaging.
We consider the GI'20 as a subdomain since all conferences are from the domain of computer science. 
As a model, we choose our \acrshort{BertLargeAMSR} model and train on two different training sets:
\begin{description}
    \item[\acrshort{nogi}] The original training dataset with all sentences from reviews of GI'20 removed.
    \item[\acrshort{withgi}] A resampling of the original training dataset of the same size as \acrshort{nogi}, with sentences from all conferences.
\end{description}
%For fair comparison we create another training set of the same size where all conferences occur with the frequency as in the original one.
% For a fair comparison, we keep the same test set and resample the training set.
% \todo{???}
% The new training set has the same size as the previous one and contains GI'20 reviews at the same fraction as our original training set. 
%We train our \acrshort{BertBaseAMSR} model, which showed better performance on smaller training sets.
Table~\ref{tab:gen_results} presents the experimental results.
% todo: comparison with table 3, stance detection seems harder on GI reviews and also argument detection and joint detection are a bit harder on GI compared to table 3.
%\todo{Although we observe a small performance decrease for argument detection, on the whole, we observe similar performance on both datasets. The more considerable performance drop on the stance task is due to the more restricted labels (only arguments). Therefore, it does not achieve the same generalization compared to the other tasks.}
We observe a small performance decrease on both tasks, about two points on argument detection and three on joint detection tasks. At the same time, we also observe similar behavior when comparing results obtained on the whole test set (Table~\ref{tab:results}) and only on GI'20 reviews by the ALL model. Therefore the more considerable drop is not necessary due to the worse generalization and can be explained by the more challenging task. Overall, the drops are relatively small, and we conclude that the model generalizes well across subdomains.

\subsection{Relevance for Decision-Making}
% \begin{table}
%     \centering
%     \caption{
%     %ToBERT Performace
%     Evaluation of acceptance classification performance in $F_1$-measure based on different sentence selection methods.
%     Using the top k\% sentences according to argumentativeness likelihood results in superior performance compared to random selection.
%     With 50\% of the text, almost the same performance is reached than with the full review.
%     \todo{Replace by figure?}
%     }
%     \label{tab:tobert}
%     \begin{tabular}{rrr}
%         \toprule
%         Size & \multicolumn{2}{c}{Selection} \\
%         & Random & Top $k$ \\
%         \midrule
%         20\% & 0.716 & 0.748 \\ 
%         30\% & 0.775 & 0.790 \\
%         40\% & 0.803 & 0.847 \\
%         50\% & 0.824 & 0.874 \\
%         \midrule
%         100\% & \multicolumn{2}{c}{0.881} \\
%         % \midrule
%         % \multirow{4}{*}{\rotatebox{90}{Top k\%}}
%         % & 20\% & 0.748 \\ 
%         % & 30\% & 0.790 \\
%         % & 40\%  & 0.847 \\
%         % & 50\% & 0.874 \\
%         % \midrule
%         % & 100\% & 0.8808 \\
%         \bottomrule
%     \end{tabular}\\
%     %\includegraphics[width=\linewidth]{figures/decision}
% \end{table}
\begin{figure}
    \centering
    \includegraphics[width=.85\linewidth]{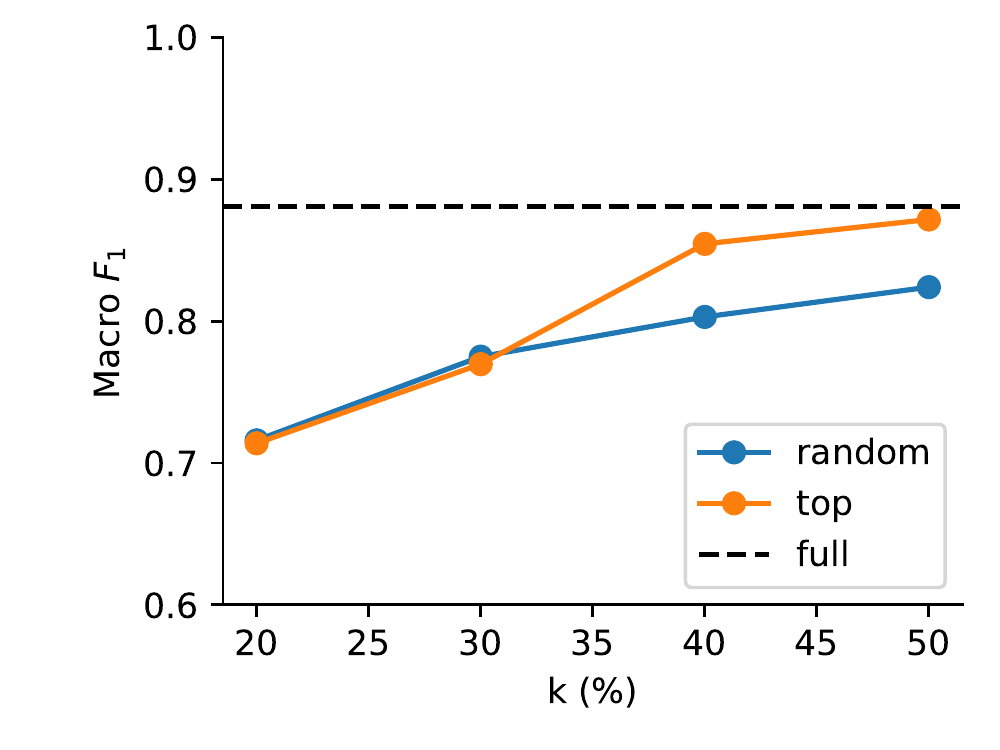}
    \caption{
    Evaluation of acceptance classification performance in $F_1$-measure based on different sentence selection methods.
    Using the top k\% sentences according to argumentativeness likelihood results in superior performance compared to random selection.
    With 50\% of the text, almost the same performance is reached as with the full review.
    }
    \label{fig:tobert}
\end{figure}
% The paper-acceptance-prediction tasks subsequently evaluates the argument extraction quality by using only the found argument in the reviews $r_1, \ldots, r_n \in R_m$ to predict for a given paper $p_m \in P$ if it should be either accepted or rejected.
In previous experiments, we have shown that peer-reviews contain arguments and these arguments can be identified automatically.
In this section, we want to verify the usefulness of the extracted arguments for the decision making process.
%As a proxy to evaluate the usefulness, we design an experiment where we aim to predict the acceptance/rejection decision solely based on the extracted arguments and compare it against the classification result using the full review.
As a proxy to evaluate the usefulness, we design an experiment where the acceptance/rejection decisions made solely by considering arguments are compared to the decisions supported by taking full reviews into account.
Therefore, we use the unannotated rest of our dataset and assign a probability to be an argument to each sentence with our best performing \acrshort{BertLargeAMSR} model.
Now, we can compare three different settings for the decision-making process:
\begin{description}
    \item[Full]
    The decision-makers are allowed to see all reviews completely. 
    This particularly includes decision suggestions often encountered in reviews that are not annotated as arguments in our dataset.
    \item[Top-K Arguments]
    The decision-makers are only allowed to see the $k\%$ sentences with the highest probability to be arguments from each review. 
    Note that the high probability to be identified as an argument does not necessarily correlate with the strength of the argument.
    \item[Random-K]
    Decision-makers are only allowed to see $k\%$ randomly selected sentences from each review. 
    We do not exclude explicit decision suggestions here.
\end{description}
We consider sentence level in this experiment despite the better performance of our model on the token-level. 
The main reason is a fair comparison with the Random-$k$ setting, random sampling of words would result in large gaps and meaningless texts, especially for small $k$.

To avoid manual expenditure, we decide to apply a language model as a decision-maker. 
Since we also have a decision for each paper in our dataset, we train models to make an acceptance/rejection decision for the different settings described above.
The standard BERT model is not directly applicable for this task since combining the reviews for a single paper often exceeds the input length restriction of at most 512 tokens.
Therefore, we employ \acrshort{ToBert}~\cite{pappagari2019hierarchical}, a model proposed for the classification of the long texts.
It splits texts into multiple segments and individual segments are first used for the finetuning of the BERT model. 
In a second step, a second transformer model on the top combines representations of the segments and makes the final decision.

The results in terms of $F_1$-measure are given in Figure~\ref{fig:tobert}.
We observe that selecting according to argumentativeness likelihood improves classification performance consistently in terms of $F_1$, compared to the random selection baseline, if at least a third of the review text is taken into consideration.
The fraction of argumentative sentences in the annotated part of our dataset is 60\%, cf. Table~\ref{tab:annotations}. 
We can achieve almost the same performance as the classifier trained on the full reviews while only considering 50\% of the review.
This is particularly impressive considering that reviews often already contain decision suggestions. 
Therefore, we conclude that arguments, which can be automatically extracted from reviews, are essential for the decision making process.

\section{Conclusion}
In this work, we have presented a new Argument Mining based approach for the assistance of different actors in the peer-review process.
We have demonstrated that arguments are present in peer-reviews and that their identification with different stances can be made automatically. 
We have also shown that the peer-review domain is different from other previous Argument Mining applications, and therefore, there is a need for a new dataset.
We have presented a new dataset that we make available for the community and have performed an extensive evaluation. 
We have also analyzed the editorial decision-making process and have empirically demonstrated that it is driven by argumentation.

In future work, we plan to address the problem of automatic determination of argument strength.
Ranking arguments, according to their strength, is an undoubtedly useful feature for the potential application. 
For this purpose, we intend to extend our decision-making model and analyze single arguments' influence on the final decision.

Another useful feature, especially for the editorial team, would be identifying similar arguments in different reviews of the same paper.

\section{Acknowledgments}
This work has been funded by the German Federal Ministry of Education and Research (BMBF) under Grant No. 01IS18036A and by the Deutsche Forschungsgemeinschaft (DFG) within the project Relational Machine Learning for Argument Validation (ReMLAV), Grant Number SE 1039/10-1, as part of the Priority Program "Robust Argumentation Machines (RATIO)" (SPP-1999).
The infrastructure for the course was provided by the Leibniz-Rechenzentrum.
The authors of this work take full responsibilities for its content.
\bibliography{references}
\end{document}

\section{Computing \& Software Infrastructure}
The experiments were conducted on a Ubuntu 18.04 system with an Intel Xeon Processor (Skylake) CPU with 20 CPU-Cores, 362 GB memory, and a Nvidia Tesla V100 GPU with 16 GB memory. 
We further used Python 3.7, PyTorch 1.4 and the Huggingface-Transformer library (2.11.0).
As a web-crawler we used the OpenReview Crawler\urlfootnote{https://openreview-py.readthedocs.io/en/latest/getting_data.html}. For the annotation of the arguments we used WebAnno \urlfootnote{https://webanno.github.io/webanno/}

\section{Hyperparameters}
\subsection{General}
In the following, we describe our general experimental setup.
For the evaluation we initialized all methods for \textbf{ten} runs with different seeds and reported the \textbf{mean Macro F$_{1}$ score}.
We used early stopping with a patience of three on a pre-selected validation set for regularization.
As loss function we either used weighted binary-cross-entropy for the Argument and Stance Detection task, or weighted cross-entropy for the Joint task.

\subsection{\acrshort{BertBaseAMSR}}
\acrshort{BertBaseAMSR} uses either a finetuned \texttt{bert-base-cased} or a \texttt{bert-large-cased} from the huggingface-transformer library.
We selected a batch size of 100 for bert-base and 32 for bert-large.
We used AdamW as an optimizer with a learning rate of $10^{-5}$ for all models.
For argument mining we selected the hyperparameters as described in the original BERT publication (Devlin et al. 2018).

\subsection{\acrshort{ToBert}}
\begin{table}
    \centering
    \begin{tabular}{ll}
    \toprule
    Hyperparameter & Values \\
    \midrule
    attention-layers  & \textbf{2}, 4, 8 \\
    attention-dimension & \textbf{512}, 1024, 2048 \\
    dropout-rates & \textbf{0.1}, 0.2, 0.4 \\
    segment-aggregations & \textbf{mean}, max, concatenation \\
    \end{tabular}
    \caption{%
    Hyperparameter values evaluated for the \acrshort{ToBert} method.
    \textbf{Bold numbers} denote the best performing ones regarding the Macro F$_{1}$ score.
    }
    \label{tab:hyperparam}
\end{table}
\acrshort{ToBert} splits long reviews in multiple segments (here we used 512 as the maximal sequence length) and the individual segments are used for finetuning of BERT.
The second step includes a transformer layer on top that combines the fine-tuned feature-representations of the the individual segments and makes the paper acceptance/rejection decision.
We evaluated multiple architectures as given in Table~\ref{tab:hyperparam}.
\textbf{Bold numbers} denote the best performing hyperparameters regarding the Macro F$_{1}$ score.

\section{Additional Experiments}
We further include the standard-deviation of some argument mining methods as a measure of variation (see Table \ref{tab:std-sentence} and Table \ref{tab:std-token})\footnote{Due to resource limitations, we could train only a single model on the AURC dataset and therefore we do not provide standard-deviations for the BertBaseAURC model. We will submit standard-deviations for the camera-ready version}.

\begin{table}
    \centering
    \begin{tabular}{lrrr}
    \toprule
    Model & Argument & Stance & Joint \\
    \midrule
    \acrshort{BertBaseAURCAMSR} & .718 $\pm$ .007 & .852 $\pm$ .016 & .734 $\pm$ .002 \\
    \acrshort{BertBaseAMSR} & .746 $\pm$ .011 & \textbf{.905} $\pm$ .023 & .734 $\pm$ .011\\
    \acrshort{BertLargeAMSR} & \textbf{.750} $\pm$ .011 & .902 $\pm$ .015 & \textbf{.738} $\pm$ .017\\
     \bottomrule
    HP &  .885 $\pm$ .056 & .978 $\pm$ .016 & .881 $\pm$ .059\\
    \bottomrule
    \end{tabular}
    \caption{
        Overview of the results for different Argument Mining tasks on \textbf{sentence level}.
        We show results in terms of mean and standard-deviations $\sigma^{2}$ of Macro $F_1$ for different BERT model variants, as well as the human performance estimate.
        In bold font we highlight the best performance of our models per task.
    }
    \label{tab:std-sentence}
\end{table}

\begin{table}
    \centering
    \begin{tabular}{lrrrrrr}
    \toprule
    Model & Argument & Stance & Joint \\
    \midrule
    \acrshort{BertBaseAURCAMSR}  & .877 $\pm$ .006 & .862 $\pm$ .016 & .796 $\pm$  .014\\
    \acrshort{BertBaseAMSR} & .886 $\pm$ .017 & .853 $\pm$ .041  & .814 $\pm$ .010\\
    \acrshort{BertLargeAMSR} & \textbf{.890} $\pm$ .012 & \textbf{.942} $\pm$ .021 & \textbf{.831} $\pm$ .014 \\
    \bottomrule
    HP  &  .873 $\pm$ .062 & .980 $\pm$ .014 & .860 $\pm$ .066\\
    \bottomrule
    \end{tabular}
    \caption{
        Overview of the results for different Argument Mining tasks on \textbf{token level}.
        We show results in terms of themean and standard-deviations $\sigma^{2}$ of Macro $F_1$ for different BERT model variants, as well as the human performance estimate.
        In bold font we highlight the best performance of our models per task.
    }
    \label{tab:std-token}
\end{table}

%todo variance